\documentclass[prl,reprint,a4paper,floatfix,amsmath,amssymb,amsfonts,noshowpacs,superscriptaddress,longbibliography]{revtex4-1}
\usepackage{newtxtext,newtxmath}
\usepackage[utf8]{inputenx}
\usepackage{graphicx}
\usepackage{booktabs}
\usepackage{mathrsfs}
\usepackage[dvipsnames]{xcolor}
\usepackage{verbatim}
\usepackage{units}
\usepackage{chemarr}
\usepackage{chemist}
\usepackage[
unicode=true,bookmarks=false,breaklinks=false,pdfborder={0 0 1},colorlinks=true]{hyperref}
\hypersetup{linkcolor=blue,citecolor=blue,urlcolor=blue}
\usepackage[textwidth=17.5cm,textheight=23.5cm,verbose,pdftex]{geometry}
\usepackage{soul}

\newcommand{\fc}{$\phi_{c}$}
\newcommand{\fr}{$\phi_{r}$}

\newcommand{\fgso}{$\phi_{\text{Ga$_2$O}}$}
\newcommand{\fiso}{$\phi_{\text{In$_2$O}}$}
\newcommand{\fsso}{$\phi_{\text{SnO}}$}
\newcommand{\fsof}{$\phi_{\text{Me$_x$O$_{y-x}$}}$}

\newcommand{\fo}{$\phi_{\text{o}}$}

\newcommand{\tg}{$T_\text{G}$}
\newcommand{\meo}{Me$_x$O$_y$}
\newcommand{\mso}{Me$_x$O$_{y-x}$}
\newcommand{\alo}{Al$_2$O$_3$}
\newcommand{\gao}{Ga$_2$O$_3$}
\newcommand{\gso}{Ga$_2$O}

\newcommand{\ino}{In$_2$O$_3$}
\newcommand{\iso}{In$_2$O}

\newcommand{\sno}{SnO$_2$}
\newcommand{\sso}{SnO}
\newcommand{\algao}{(Al$_x$Ga$_{1-x}$)$_2$O$_3$}

\newcommand{\ingao}{(In$_x$Ga$_{1-x}$)$_2$O$_3$}
\newcommand{\pfu}{$\,\text{nm}^{-2}\,\text{s}^{-1}$}
\newcommand{\utg}{$\,^{\circ}\text{C}$}

\newcommand{\gr}{\Gamma}

\begin{document}


\graphicspath{{./figs/}}

\title{Extending the kinetic and thermodynamic limits of molecular-beam epitaxy utilizing suboxide sources or metal-oxide catalyzed epitaxy}

\author{Patrick Vogt}
\email[Electronic mail: ]{pv269@cornell.edu}
\affiliation{Department of Materials Science and Engineering, Cornell University, Ithaca 14853, New York, USA}
\author{Felix V.~E.~Hensling}
\affiliation{Department of Materials Science and Engineering, Cornell University, Ithaca 14853, New York, USA}
\author{Kathy Azizie}
\affiliation{Department of Materials Science and Engineering, Cornell University, Ithaca 14853, New York, USA}
\author{Jonathan P.~McCandless}
\affiliation{School of Electrical and Computer Engineering, Cornell University, Ithaca 14853, New York, USA}
\author{Jisung Park}
\affiliation{Department of Materials Science and Engineering, Cornell University, Ithaca 14853, New York, USA}
\author{Kursti DeLello}
\affiliation{School of Applied and Engineering Physics, Cornell University, Ithaca 14853, New York, USA}
\author{David A.~Muller}
\affiliation{School of Applied and Engineering Physics, Cornell University, Ithaca 14853, New York, USA}
\affiliation{Kavli Institute at Cornell for Nanoscale Science, Ithaca 14853, New York, USA}
\author{Huili G.~Xing}
\affiliation{Department of Materials Science and Engineering, Cornell University, Ithaca 14853, New York, USA}
\affiliation{School of Electrical and Computer Engineering, Cornell University, Ithaca 14853, New York, USA}
\affiliation{Kavli Institute at Cornell for Nanoscale Science, Ithaca 14853, New York, USA}
\author{Debdeep Jena}
\affiliation{Department of Materials Science and Engineering, Cornell University, Ithaca 14853, New York, USA}
\affiliation{School of Electrical and Computer Engineering, Cornell University, Ithaca 14853, New York, USA}
\affiliation{Kavli Institute at Cornell for Nanoscale Science, Ithaca 14853, New York, USA}
\author{Darrell G.~Schlom}
\email[Electronic mail: ]{schlom@cornell.edu}
\affiliation{Department of Materials Science and Engineering, Cornell University, Ithaca 14853, New York, USA}
\affiliation{Kavli Institute at Cornell for Nanoscale Science, Ithaca 14853, New York, USA}
\affiliation{Leibniz-Institut f{\"u}r Kristallz{\"u}chtung, Max-Born-Str.~2, 12489 Berlin, Germany}

\begin{abstract}
We observe a catalytic mechanism during the growth of III-O and IV-O materials by suboxide molecular-beam epitaxy ($S$-MBE). By supplying the molecular catalysts \iso\ and \sso\ we increase the growth rates of \gao\ and \ino. This catalytic action is explained by a metastable adlayer $A$, which increases the reaction probability of the reactants \gso\ and \iso\ with active atomic oxygen, leading to an increase of the growth rates of \gao\ and \ino. We derive a model for the growth of binary III-O and IV-O materials by $S$-MBE and apply these findings to a generalized catalytic description for metal-oxide catalyzed epitaxy (MOCATAXY), applicable to elemental and molecular catalysts. We derive a mathematical description of $S$-MBE and MOCATAXY providing a computational framework to set growth parameters in previously inaccessible kinetic and thermodynamic growth regimes when using the aforementioned catalysis. Our results indicate MOCATAXY takes place with a suboxide catalyst rather than with an elemental catalyst. As a result of the growth regimes achieved, we demonstrate a \gao/\alo\ heterostructure with an unrivaled crystalline quality, paving the way to the preparation of oxide device structures with unprecedented perfection.
\end{abstract}


\maketitle
\section{I.~Introduction}
\noindent
Molecular-beam epitaxy (MBE) takes place under non-equilibrium conditions and surface kinetics plays a dominant role in the MBE growth process---allowing growth modes to be intentionally manipulated \cite{arthur1968,copel1989,neugebauer2003,lewis2017,vogt2017a}. For decades, the single-step reaction mechanism occurring during the MBE growth of III-V (e.g., GaAs, GaN, AlN) \cite{ploog1982,calleja1999,garrido2008} and II-VI (e.g., ZnSe, ZnO) \cite{zhu1989,kato2003} compound semiconductors has defined MBE as a rather simple and straightforward thin film technique, especially when compared with chemical vapor deposition methods \cite{hirako2005}. In contrast to the growth of III-V and II-VI materials, the surface kinetics of III-O (e.g., \gao\ and \ino) \cite{tsai2010,vogt2015a,vogt2016a,vogt2016b,vogt2016c,vogtdiss,vogt2018a} and IV-O (e.g., \sno) \cite{white2008} compounds is governed by a complex reaction pathway, resulting in a two-step reaction mechanism to form the intended compound. The formation and subsequent desorption of ad-molecules, called suboxides (e.g., \gso, \iso, \sso), define the growth-limiting step for these classes of materials. The result is a rather narrow growth window within the adsorption-controlled regime \cite{white2008,vogt2016a,vogt2016b,oshima2018,vogt2018a}.

By engineering the MBE processes, two variants of MBE have recently been developed. These variants extend the kinetic and thermodynamic limits within which III-O and IV-O materials may be grown. The first variant, \textit{suboxide} MBE ($S$-MBE) \cite{vogt2021}, refers to a technique that uses the decomposition of III-VI and IV-VI compounds (i.e., \meo) by group III and IV elements \cite{vogt2015a,vogtdiss} as well as a special MBE source chemistry \cite{vogt2021} to produce suboxide molecular beams that consist almost entirely (typically > 99.9\%) of a single suboxide molecular species (i.e., \mso). Stoichiometric coefficients of \meo\ and \mso\ are $x = 2$ and $y = 3$ for III-O (e.g., \gao, \ino) and $x = 1$ and $y = 2$ for IV-O (e.g., \sno) materials.

Using the $S$-MBE approach, the growth-limiting step occurring during the growth of III-O and IV-O materials by conventional MBE is bypassed, enabling the growth of films with excellent structural quality and surface smoothness at growth rates exceeding $1\,\upmu\text{m}\,\text{hr}^{-1}$ and comparatively low growth temperature (\tg) \cite{vogt2021}. The second variant is metal-oxide catalyzed epitaxy (MOCATAXY), a method involving the introduction of a catalyst into oxide growth systems \cite{vogtdiss,vogt2017a,kracht2017,vogt2018a}. Using this technique, the growth-limiting step of III-O compounds is bypassed by the catalyst \cite{vogt2017a}. It has been proposed that MOCATAXY results from metal-exchange catalysis (MEXCAT) \cite{vogt2017a} and has been observed on various growth surfaces for the formation of \gao\ and \algao\ \cite{vogt2017a,kracht2017,vogt2018a,mazzo2019,mauze2020,mauze2020a} as well as during other physical vapor deposition methods \cite{kneiss2019}. Nevertheless, the underlying physics leading to the observed catalysis has been explained differently---experimentally \cite{vogt2017a,kracht2017} and theoretically \cite{wang2020}---and thus remains disputable.
\begin{figure}[t]
\includegraphics[scale=0.915]{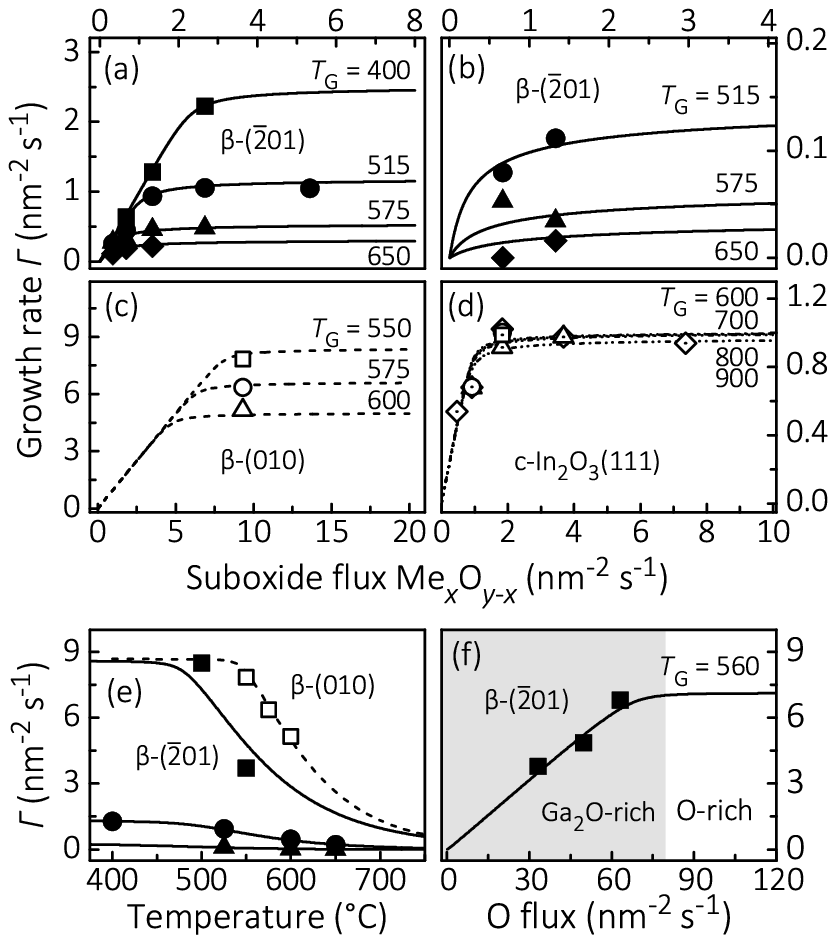}
\caption{(a) and (b) the dependence of the film growth rate (\gr, in units of the flux of \gso\ being incorporated into the growing \gao\ film) of $\upbeta$-\gao($\bar{2}$01) by $S$-MBE on the flux of the suboxide \gso\ at active atomic oxygen fluxes $\text{\fo} = 5.7\text{\pfu}$ and $\text{\fo} = 0.7\text{\pfu}$, respectively. (c) \gr\ of $\upbeta$-\gao(010) at $\text{\fo} = 33\text{\pfu}$ as a function of \fgso. (d) Dependence of \gr\ of bixbyite \ino(111) (in units of the flux of \iso\ being incorporated into the growing \ino\ film) on the flux of the suboxide \iso, at $\text{\fo} = 2.0\text{\pfu}$. (e) Growth temperature (\tg) dependence of \gr\ of $\upbeta$-\gao($\bar{2}$01) and $\upbeta$-\gao(010) at different active atomic oxygen fluxes:~\fo\ = 28.5\text{\pfu} (solid and hollow squares), \fo\ = 5.7\text{\pfu} (solid circles), and \fo\ = 0.7\text{\pfu} (solid triangles). The flux of \gso\ supplied, \fgso, corresponds to the vertical axis intercepts at low \tg. (f) \gr\ of $\upbeta$-\gao($\bar{2}$01) as a function of \fo\ at a constant value of \fgso\ given by the vertical-axis intercept at high \fo. Gray and white areas in (f) indicate the \gso-rich and O-rich flux regimes, respectively. \tg\ in \utg\ is indicated in the figures. Symbols are experimental data and lines are model predictions according to Eqs.~(\ref{eq:m1})--(\ref{eq:kappa}) and the parameters given in Table \ref{tab} for the $S$-MBE growth of $\upbeta$-\gao($\bar{2}$01) [solid symbols and lines], $\upbeta$-\gao(010) [hollow symbols and dashed lines], and bixbyite \ino(111) [hollow symbols with a dot in their center and dashed-dotted lines]. Data in (a) are taken from Ref.~\cite{vogt2021}}.
\label{fig:ino}
\end{figure} 

In this Letter, the combination of $S$-MBE and MOCATAXY is investigated and shown to result in a marked extension of the kinetic and thermodynamic limits of the growth of \gao\ and \ino. By supplying suboxide molecular beams of \gso, \iso, and \sso, a growth rate (\gr) enhancement of \gao\ and \ino\ is observed. Quantitative models describing this enhanced growth rate during $S$-MBE as well as during the catalyzed $S$-MBE of III-O thin films are derived and applied to the experimental observations to extract model parameters. Through a systematic comparison of experimental \gr\ data from different growth systems and growth methods (i.e., MOCATAXY during $S$-MBE and conventional MBE), a generalization of the proposed MEXCAT mechanism \cite{vogt2017a} is developed. This generalized growth mechanism is applicable to the growth of \gao\ and \ino\ by conventional MBE and $S$-MBE. 
\section{II.~Suboxide MBE ($S$-MBE) model}
\noindent
We begin by deriving a growth-rate model for the growth of III-VI and IV-VI compounds with general formula \meo\ by $S$-MBE and validate it by using \gao\ and \ino\ as examples. Figure \ref{fig:ino} shows the growth kinetics of \gao\ and \ino\ as a function of their respective growth parameters. In panels (a)--(d), the film growth rate, \gr, is observed to increase linearly with the incident flux of suboxide, \fsof\ (i.e., \fgso\ or \fiso), in the O-rich regime. This linear increase in \gr\ reaches a plateau in the adsorption-controlled regime when \fsof\ exceeds the flux of active atomic oxygen, \fo. The growth kinetics for the growth of III-O compounds by $S$-MBE are thus the same as those of III-V and II-VI materials when grown by conventional MBE \cite{vogt2021}. 

Figures \ref{fig:ino}(e) and \ref{fig:ino}(f) depict \gr\ as a function of \tg\ and \fo, respectively. For the same growth conditions, \gr\ of \gao\ on \gao(010) [hollow squares] is larger than that of \gao\ on \gao($\bar{2}$01) [solid squares]; see panel (e). This result is similar to the growth of \gao\ by conventional MBE on \gao($\bar{2}$01) versus on \gao(010) substrates \cite{sasaki2012,oshima2018,mazzo2020}. Comparing the growth kinetics of \gao($\bar{2}$01) and \ino(111) [e.g., the data in panel (a) with the data in panel (d)] establishes that the range of \tg\ within which high-quality films of \ino\ can be grown at high \gr\ is larger than that for \gao. This result is also similar to the growth of \gao\ and \ino\ by conventional MBE \cite{vogt2015a,vogt2016b,vogt2018a}. 

\begin{figure}[b]
\includegraphics[scale=0.9]{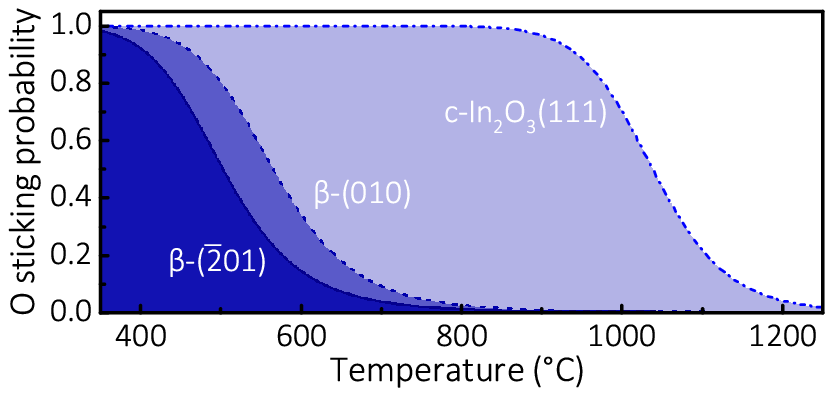}
\caption{Oxygen sticking probability $\varsigma$ as a function of the growth temperature \tg. Solid, dashed, and dashed-dotted lines represent $\varsigma$ of $\upbeta$-$\text{\gao}(\bar{2}01)$ (the dark blue area), $\upbeta$-$\text{\gao}(010)$ (the blue area), and bixbyite (cubic) $\text{c-\ino}(111)$ (the pale blue area), respectively. The lines are modeled by Eq.~(\ref{eq:varsig}) using the parameters given in Table \ref{tab} for the respective phases.}
\label{fig:stick}
\end{figure}

To model the growth of binary oxides (\meo) from their suboxides (\mso), we take the single-step reaction kinetics of $S$-MBE \cite{vogt2021} into account. Here, the growth takes place via the reaction
\begin{align}
\label{eq:fo}
\text{Me}_x\text{O}_{y-x}\,(a) + x \text{O}\,(a) \reactrarrow{0pt}{1cm}{$\kappa$}{} \text{Me}_x\text{O}_y\,(s) \,, 
\end{align} 
with constant $\kappa$ describing the \meo\ formation rate. Adsorbate and solid phases are denoted as $a$ and $s$, respectively. Based on reaction (\ref{eq:fo}), we set up a generalized growth-rate model describing the growth of these materials by $S$-MBE:
\begin{align}
\frac{\text{d}n_{r}}{\text{d}t} &= \phi_{r} - \kappa n_{r} n_{\text{o}}^x - \tau_{r}^{-1}  n_{r} \,,\label{eq:m1}
\\
\frac{\text{d}n_{\text{o}}}{\text{d}t} &= \varsigma  \phi_{\text{o}} - x \kappa  n_{r}  n_{\text{o}}^x - \tau_{\text{o}}^{-1} n_{\text{o}} \,,\label{eq:m2}
\\
\frac{\text{d}n_{p}}{\text{d}t} &= \Gamma = \kappa n_{r} n_{\text{o}}^x \,.\label{eq:m3}
\end{align}
The surface densities of adsorbed cationic reactants ($r = \text{\mso}$), O adsorbates, and formed products ($p = \text{\meo}$) are denoted as $n_r$, $n_\text{o}$, and $n_p$, respectively. Their time derivative is described by the operator $\text{d}/\text{d}t$. The surface lifetimes of $r$ and O are given by $\tau_{r}$ and $\tau_{\text{o}}$, respectively. The \tg-, orientation-, and growth-system-dependent sticking probability $\varsigma$ of O can be derived by considering two competing processes:~the chemisorption of O and the desorption of O from the given adsorption sites, described by the reaction rates $k_{\text{c}}$ and $k_{\text{d}}$, respectively,
\begin{equation}
\label{eq:sk}
k_i = \nu_i  e^{-\frac{\varepsilon_i}{k_{\text{B}}\,\text{\tg}}} \,,
\end{equation}
with $i = \text{c,\,d}$, frequency factor $\nu_i$, and the respective energy barrier $\varepsilon_i$. Following the Kisliuk model \cite{kis1957},  the O sticking probability $0 \leq \varsigma \leq 1$ \cite{lombardo1991,lipponer2012} is given by
\begin{equation}
\label{eq:varsig}
\varsigma = \frac{k_{\text{c}}}{k_{\text{c}} + k_{\text{d}}} = \left(1 + \nu e^{-\frac{\updelta\varepsilon}{k_{\text{B}}T_{\text{G}}}}\right)^{-1}
\end{equation}
with dimensionless pre-factor $\nu = \nu_{\text{d}} / \nu_{\text{c}}$ and activation barrier $\updelta\varepsilon = \varepsilon_{\text{d}} - \varepsilon_{\text{c}}$. In the case of a high O desorption barrier or a high O adsorption barrier, the limits of Eq.~(\ref{eq:varsig}) are (i) $\varepsilon_{\text{d}} \gg \varepsilon_{\text{c}} \Rightarrow \updelta\varepsilon \gg 0 \Rightarrow \varsigma \rightarrow 1$ and (ii) $\varepsilon_{\text{c}} \gg \varepsilon_{\text{d}} \Rightarrow \updelta\varepsilon \ll 0 \Rightarrow \varsigma \rightarrow 0$, respectively, satisfying our conditions of $0 \leq \varsigma \leq 1$. Figure \ref{fig:stick} depicts the model results of Eq.~(\ref{eq:varsig}) of our model for \gao\ and \ino\ using the parameters given in Table \ref{tab}.
\begin{table}[t]
  \centering	
    \begin{tabular}{||c|c|c|c|c||}
		 \toprule
      &$\; \nu$\;  &$\; \updelta\varepsilon \, (\text{eV}$)\; &$\; P_0 \, \bigl(\text{nm}^{2x}\bigr) \;$ &$\; \mathcal{E} \, (\text{eV}$)\;  \\ 
		 \midrule
	  	$\; \upbeta$-\gao($\bar{2}$01) \; & $e^{16.1 \pm 1}$  & $1.07 \pm 0.1$  & $e^{0.15 \pm 0.05}$  & $0.12 \pm 0.01$    \\ \hline
			$\; \upbeta$-\gao(010) \;         & $e^{16.6 \pm 1}$  & $1.20 \pm 0.2$  & $e^{0.25 \pm 0.09}$  & $0.15 \pm 0.03$    \\ \hline
			$\;$\ino(111) \;                  & $e^{28.4 \pm 1}$  & $3.21 \pm 0.5$  & $e^{0.90 \pm 0.10}$  & $0.25 \pm 0.02$    \\ \hline
\bottomrule
	\end{tabular} 
\caption{Parameters $\nu$, $\updelta\varepsilon$, $P_0$, and $\mathcal{E}$ obtained using an iterative approach of the growth model to the flux-dependence of the growth rate (\gr) of \gao($\bar{2}$01), \gao(010), and \ino(111) by $S$-MBE with Eqs.~(\ref{eq:m1})--(\ref{eq:kappa}).}\label{tab}
\end{table}	
 
Solving Eqs.~(\ref{eq:m1}) and (\ref{eq:m2}) with respect to $n_r$ and $n_{\text{o}}$ and inserting their solutions into Eq.~(\ref{eq:m3}) yields an analytical expression for \gr\ described by three kinetic parameters: $\kappa$, $\tau_r$, and $\tau_{\text{o}}$, see Eqs.~(\ref{eq:s1})--(\ref{eq:s7}).
To reduce the complexity of the model we assume that O adsorbate desorption is negligible and use $\kappa \gg \tau_{\text{o}}^{-1}$ and $\tau_r \ll \tau_{\text{o}}$. Assuming that the observed reaction processes follow a thermally activated Arrhenius behavior, we further reduce the complexity of our model by forming the product
\begin{equation}
\label{eq:kappa}
P\bigl(\text{\tg}\bigr) = \tau_r \kappa  =  P_0 \, e^{\left(\frac{\mathcal{E}}{k_{\text{B}}\,\text{\tg}}\right)} \,,
\end{equation}
with pre-exponential factor $P_0 = \tau_r^0 \kappa_0$ and activation energy $\mathcal{E} = \varepsilon_{\tau_r} - \varepsilon_{\kappa}$. The latter expression yields a larger $\kappa$ or a longer $\tau_r$ the higher the growth rate \gr\ of the intended compound. This is in agreement with experiment as shown in Fig.~\ref{fig:ino}.

We apply the solution of this model to the binary growth rate data of \gao\ and \ino\ by $S$-MBE depicted in Fig.~\ref{fig:ino} and extract the kinetic parameters summarized in Table \ref{tab}. To extract the parameters we use an iterative approach. For example, for the growth of \gao\ by $S$-MBE we first establish that the functional form of the equations accurately describe the growth of ($\bar{2}$01) oriented $\upbeta$-\gao\ films, as plotted in Figs.~\ref{fig:ino}(a) and \ref{fig:ino}(b).
Having established that the functional form of the equations [solutions given by Eqs.~(\ref{eq:s1})--(\ref{eq:s7})] accurately
describe the growth of ($\bar{2}$01) oriented $\upbeta$-\gao\ films, the model is next expanded to an additional orientation, (010) $\upbeta$-\gao\
films using the data in Fig.~\ref{fig:ino}(c).
\subsection{Orientation-dependent growth rate of $\upbeta$-\gao}
As a quantitative result, we find that the range of \tg\ that can be used to produce high-quality epitaxial films at high \gr\ (i.e., a growth window) of ($\bar{2}$01)-oriented \gao\ films is narrower than for the growth of (010)-oriented \gao\ films, which, in turn, is narrower than the one for the growth of (111)-oriented \ino\ films. This is because of the different reaction efficiencies, $\eta$, of \gso\ and \iso\ with active O species (more detailed explanation below in the text). In addition, we explain the different sticking probabilities $\varsigma$, e.g., as obtained for \gao($\bar{2}01$) and \gao(010), by the activity of surface reactions between \gso\ and O adsorbates  being dependent on the orientation of the surface on which the reaction takes place. In other words, the reservoir of active atomic oxygen for \gso\ oxidation on \gao(010) is larger than the one on \gao($\bar{2}$01). This finding can be transferred to the growth of \gao\ by conventional MBE where a similar dependence of \gao\ growth rate is observed. We propose that this \gr\ dependence results from an orientation-dependent activity of the oxidation of \gso\ to \gao\ \cite{vogt2016a,oshima2018} and that the underlying reason is due to the orientation-dependent vertical and lateral bond strengths between ad-atoms and the substrate surfaces \cite{sasaki2012,mazzo2020}. In Refs.~\cite{sasaki2012,oshima2018,mazzo2020} the orientation dependence of the growth rate on the ($hkl$) plane, $\Gamma_{(hkl)}$, is given for $\upbeta$-\gao:~$\Gamma_{(010)} > \Gamma_{(001)} > \Gamma_{(\bar{2}01)} > \Gamma_{(100)}$. We observe this same order of growth rate as a function of surface orientation for the growth of \gao\ by $S$-MBE, and conclude that the orientation-dependent values of the sticking probability $\varsigma_{(010)} > \varsigma_{(001)} > \varsigma_{(\bar{2}01)} > \varsigma_{(100)}$ can be related to the orientation-dependent reaction activities of adorbates, underlying the observed order of $\Gamma_{(hkl)}$ for $\upbeta$-\gao, see Fig.~\ref{fig:stick}.
\section{III.~MOCATAXY combined with $S$-MBE}
\begin{figure}[t]
\includegraphics[scale=0.925]{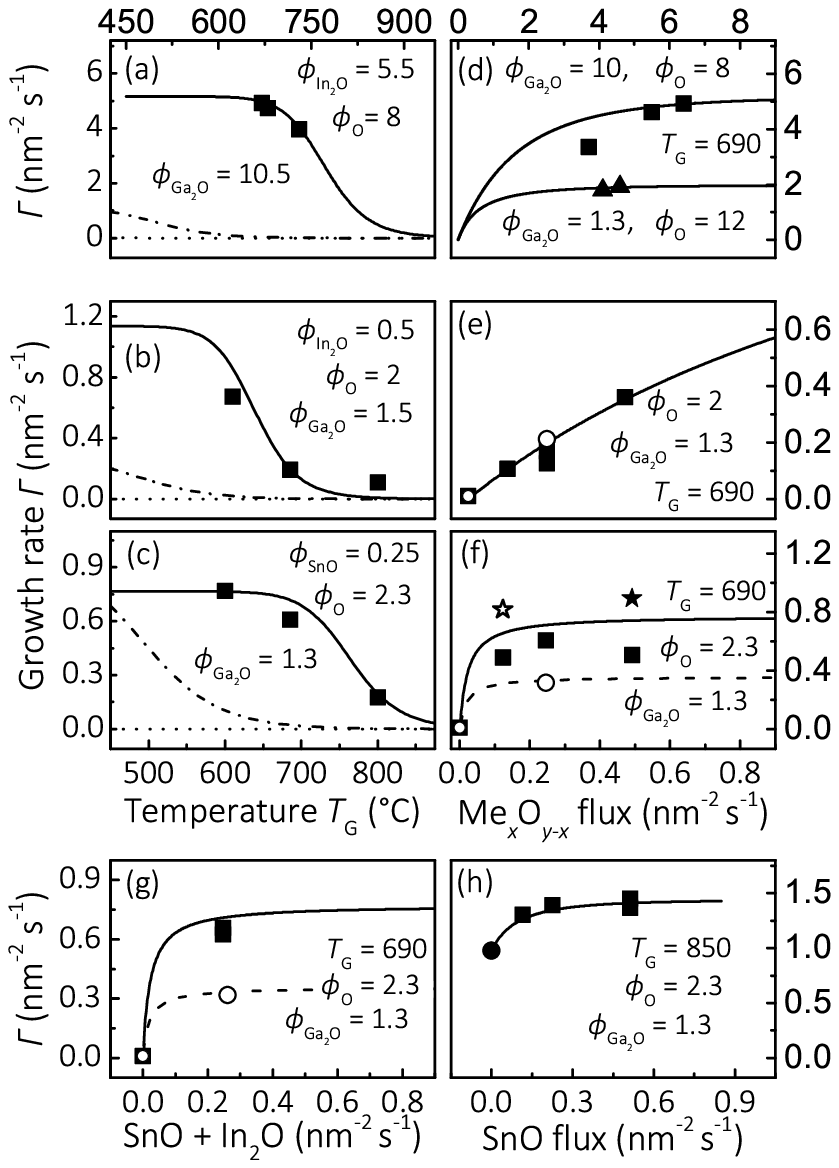}
\caption{(a)--(c) \gr\ of $\upbeta$-\gao($\bar{2}$01) as a function of \tg\ in the presence of either the catalyst \iso\ or \sso\ at the steady state fluxes given. (d)--(g) \gr\ of $\upbeta$-\gao($\bar{2}$01) (solid symbols) and $\upbeta$-\gao(010) (hollow symbols) as a function of the \iso\ flux as a catalyst [panels (d) and (e)], \sso\ flux as a catalyst [panel (f)], and their sum $a \, \text{\iso} + (1-a)\, \text{\sso}$ as competing catalysts (with $a = 0.5$) [panel (g)]. (h) \gr\ of bixbyite \ino(111) using \sso\ as a catalyst. Solid lines are model predictions by the catalytic model, Eqs.~(\ref{eq:mc1})--(\ref{eq:alpha}). The parameters used for our model predictions are given in Table \ref{tab2}. Dashed-dotted and dotted lines depict the model predictions for the growth of $\upbeta$-\gao\ by $S$-MBE [Eqs.~(\ref{eq:m1})--(\ref{eq:kappa})] and conventional MBE \cite{vogt2018b} in the absence of a catalyst, respectively. The units of the fluxes given in the figures are $\text{nm}^{-2}\,\text{s}^{-1}$; the temperature \tg\ are in \utg. The values of \fo\ correspond to the maximum available active atomic oxygen provided by the catalyst (i.e., \iso\ or \sso). Additional characterizations of the sample grown at the conditions depicted by the solid star in (f) are provided in Fig.~\ref{fig:xrd}.}
\label{fig:cat}
\end{figure}

\noindent
We next describe the enhancement to the growth rate of \gao\ and \ino\ due to the presence of the catalysts \iso\ and \sso, i.e., the combined effects of $S$-MBE and MOCATAXY. As depicted in Figs.~\ref{fig:cat}(a)--\ref{fig:cat}(c), we observe a drastic enhancement in \gr\ of heteroepitaxial $\upbeta$-\gao($\bar{2}$01) grown on \alo(0001) and homoepitaxial $\upbeta$-\gao(010) grown on $\upbeta$-\gao(010) using the suboxides \iso\ and \sso\ as catalysts. The catalytic effect of \sso\ on the growth rate of \gao\ is stronger than that of \iso. It is also stronger for the growth of $\upbeta$-\gao($\bar{2}$01) (solid symbols) than for the growth of $\upbeta$-\gao(010) (hollow symbols). The stronger catalytic effect of \sso\ compared with \iso\ can be explained by their different vapor pressures, i.e., $P_{\text{\sso}} < P_{\text{\iso}}$ \cite{colin1965,val1977,adkison2020}. Thus, under similar growth conditions, the surface lifetime of \sso\ is longer than that of \iso. Hence, \sso\ can be re-oxidized more often than \iso\ \cite{vogtdiss,vogt2017a}, and the O reservoir that ultimately ends up oxidizing \gso\ in the presence of \sso\ is larger than that for \gso\ oxidation in the presence of \iso. Figures \ref{fig:cat}(d)--\ref{fig:cat}(f) depict \gr\ of \gao\ as a function of \fiso\ and \fsso, respectively. When \sso\ is used as a catalyst it is evident from the data that the catalytic effect on \gao($\bar{2}$01) is stronger than it is on \gao(010). As shown below in Eq.~(\ref{eq:alpha}), the catalytic activity, $\alpha$, decreases with increasing O flux. Therefore, the weaker catalytic effect observed on $\upbeta$-\gao(010) compared with the effect on $\upbeta$-\gao($\bar{2}$01) may be explained by the higher surface-dependent sticking probability $\varsigma_{(010)} > \varsigma_{(\bar{2}01)}$ (values are given in Table \ref{tab}). Further studies on the atomic incorporation and surface segregation of residual Sn and In in the grown thin films, when using \sso\ and \iso\ as catalysts, respectively, need to be performed. Nevertheless, energy-dispersive x-ray spectroscopy (EDXS), secondary-ion mass spectrometry (SIMS), and atomic probe tomography (APT) have already shown that the concentrations of In \cite{vogt2017a,vogt2018a,mauze2020} and Sn \cite{kracht2017} in \gao\ as well as in \algao\ films are below $1\,\%$. This is consistent with our x-ray diffraction (XRD) results showing only diffraction peaks from \gao\ in the thin films, an example of which is shown in Fig.~\ref{fig:xrd}.

Comparing Figs.~\ref{fig:cat}(f) and \ref{fig:cat}(g) reveals that the growth rates \gr\ of $\upbeta$-\gao($\bar{2}$01) and $\upbeta$-\gao(010) using solely \sso\ and $\text{\sso} + \text{\iso}$ as catalysts are very similar (under otherwise identical growth conditions). We thus conclude that \sso\ suppresses \iso\ as a catalyst, i.e., the catalytic effects are not additive:~the presence of SnO inhibits (or isolates) the catalytic activity of \iso\ in the \sso--\iso--\gso--O system. We predict the same effect for MOCATAXY by conventional MBE, i.e., when using the elemental catalysts In \cite{vogt2017a} and Sn \cite{kracht2017} in combination for the growth of \gao\ in the Sn--In--Ga-O system.

Figure~\ref{fig:cat}(h) shows the catalytic effect of \sso\ on \gr\ of \ino(111) grown on \alo(0001) substrates. The solid circle in Fig.~\ref{fig:cat}(h) corresponds to the growth rate of \ino\ determined by x-ray reflectivity (XRR) in the absence of \sso. In the presence of \sso, the growth rate of \ino\ (again measured by XRR) is seen to increase by a factor of $\sim 1.4$, i.e., far larger than the effect caused by additional SnO incorporation into \ino. SIMS \cite{eag} and x-ray fluorescence (XRF) \cite{eag} were used to quantify the Sn content present in the \ino\ films grown with the SnO catalyst. A value of $\approx 3 \, \%$ Sn in \ino\ was observed by SIMS and XRF independent of the SnO fluxes in the $(0.2 - 0.5)\,\text{nm}^{-2}\,\text{s}^{-1}$ range.

This result indicates that \sso\ increases the available O reservoir for the oxidation of \iso\ to \ino\ by a factor of $\sim 1.4$ (at these growth conditions). This finding is in line with the data plotted in Figs.~\ref{fig:cat}(f) and \ref{fig:cat}(g), i.e., that \sso\ has a higher reactivity with O than \iso. We explain the catalytic action of \sso\ on \iso\ by their different vapor pressures $P_{\text{\sso}} < P_{\text{\iso}}$ \cite{colin1965,val1977,adkison2020}. Thus, \sso\ provides a larger O reservoir for \iso\ oxidation than is available in the absence of \sso. The \iso-\sso-O system is the first catalytic system observed for MBE growth beyond \gao-based systems \cite{vogt2017a,kracht2017,vogt2018a,mazzo2019,mauze2020,mauze2020a}. The discovery of MOCATAXY during $S$-MBE and its extension to \ino\ suggests its universality for the MBE growth of a multitude of oxide compounds (during $S$-MBE as well as conventional MBE).

We explain the observed catalysis during the growth of \gao\ and \ino\ by $S$-MBE by the formation of a metastable adlayer $A$ between the catalyst $c$ and O, e.g., $A = \text{\iso--O}$ or $A = \text{\sso--O}$. As experimentally and mathematically shown by Fig.~\ref{fig:ino}, Table \ref{tab}, and Eqs.~(\ref{eq:m1})--(\ref{eq:varsig}), \iso\ possesses a higher surface reactivity with active O than does \gso, leading to a higher growth rate of \ino\ compared with \gao\ at comparable growth conditions. This behavior is similar to the growth of \gao\ and \ino\ by conventional MBE \cite{vogt2015a,vogt2016c}, which was explained by the different oxidation efficiencies $\eta$ of the elements Sn, In, and Ga following the order:~$\eta_{\text{Sn}} \approx 1.1\,\eta_{\text{In}} \approx 3.1\,\eta_{\text{Ga}}$ \cite{vogt2015a,vogtdiss,vogt2017a}. Taking the ratio of maximum available O for \gso\ and \iso\ oxidation (data plotted in Figs.~\ref{fig:ino}(b) and \ref{fig:ino}(d) and flux conversions given in Eqs.~(\ref{eq:fc}) from mixtures of O$_2$ and 10\% O$_3$ \cite{theis1997} at a background pressure of $1 \times 10^{-6}\,\text{Torr}$, we obtain $\eta_{\text{\iso}} \approx 2.8\,\eta_{\text{\gso}}$. This result is very similar to $\eta_{\text{In}} \approx 2.8\, \eta_{\text{Ga}}$ \cite{vogt2015a,vogtdiss,vogt2017a} observed during conventional MBE growth. We surmise following Ref.~\cite{vogtdiss} that the same value of $\eta$ observed during MOCATAXY in conventional MBE and $S$-MBE arises from what is the second reaction step of conventional MBE and the sole reaction step of $S$-MBE, where the suboxide reacts with O to complete the formation of the intended oxide (e.g., $\text{\gso} \rightarrow \text{\gao}$) \cite{vogt2018a} and \textit{not} to the suboxide formation step (e.g., $\text{Ga} \rightarrow \text{\gso}$) \cite{vogt2018a}. Thus, for the discussion and analysis that follow we use $\eta_{\text{\sso}} \approx 1.1\,\eta_{\text{\iso}} \approx 3.1\,\eta_{\text{\gso}}$ during $S$-MBE. These oxidation efficiencies follow the same order in $\eta$ as was observed for growth by conventional MBE \cite{vogtdiss,vogt2017a}.
\subsection{A.~Generalized metal-exchange catalytic model}
\noindent
We propose that the role of the catalyst $c$ (e.g., \iso, \sso, In \cite{vogt2017a}, or Sn \cite{kracht2017}) is to increase the O adsorbate reservoir of the reactant $r$ (e.g., \gso\ or Ga) by forming $A$ through the reaction
\begin{equation}
\label{eq:c1}
c\,(a) + \text{O}\,(a) \xrightarrow{} A \, (a)  \, ,
\end{equation}
with examples of $A$ being $\text{\iso--O}$, $\text{In--O}$, $\text{\sso--O}$, or $\text{Sn--O}$. In the presence of $r$, $A$ is unstable and catalyzes the incorporation of $r$ into the intended product $p$ (e.g, \gao), while decreasing the reaction barrier of $r$ with O. Thus, reaction (\ref{eq:c1}) is subsequently followed by the reaction
\begin{equation}
\label{eq:c2}
\text{\mso}\,(a) + x A\,(a) \xrightarrow{} \text{\meo}\,(s) + c\,(a) \, .
\end{equation} 
Equation (\ref{eq:c2}) describes the consumption of $A$ while forming the product $p = \text{\meo}$) and releasing $c$ on the growth surface. The catalyst $c$ may be re-oxidized \cite{vogt2017a}, leading to an increase in the available O reservoir for the reactant ($r = \text{\mso}$), and thus to an extension of the kinetic and thermodynamic limits to the formation of $p$. 

An adlayer formed by In has also been observed during the formation of GaN using In as a surface active agent (surfactant) \cite{neugebauer2003}. Here, the In adlayer enables an enhanced diffusion channel for the Ga and N adsorbates. Moreover, a surface instability of In--O bonds in the presence of Fe \cite{wagner2016} and Ga \cite{vogt2017a} has been observed during conventional MBE. We emphasize, however, the catalytic effect we are describing must not be confused with effects resulting from surfactants during the growth of III-V compounds by conventional MBE \cite{copel1989,neugebauer2003,lewis2017}.

In order to describe MOCATAXY for elemental (e.g., In \cite{vogt2017a} and Sn \cite{kracht2017}) and molecular catalysts (e.g., \iso\ and \sso) as well as for different materials (e.g., \gao\ and \ino) mathematically, we would have to take into account the surface populations of $c$, $r$, and atomic O together with the surface density of the $p$ that forms following the Langmuir-Hinshelwood mechanism \cite{wint1997}. We may reduce the complexity of the model significantly by only taking into account the most likely reactions involved in the formation of $p$, following the Eley-Rideal formalism \cite{eley1940}. The resulting set of coupled differential equations reads:
\begin{align}
\frac{dn_{c}}{dt} &= \phi_{c} - \sigma n_{c} (1-\theta_A) \text{\fo} + \alpha \text{\fr} \theta_{A}^x  - \gamma_{c} n_{c} \,,\label{eq:mc1}
\\
\frac{d\theta_{A}}{dt} &= \sigma n_{c} (1-\theta_{A}) \text{\fo} - \alpha \text{\fr} \theta_{A}^x  \,,\label{eq:mc2}
\\
\frac{dn_{p}}{dt} &\equiv \text{\gr} = \alpha \text{\fr} \theta_{A}^x  \,,\label{eq:mc3}
\end{align}
with the adatom density $n_{c}$ and desorption rate constant $\gamma_c$ of the catalyst $c$, as well as the surface coverage $\theta_{A}$ of $A$. The second and third terms in Eq.~(\ref{eq:mc1}) refer to the formation rates of $A$ and $p$, respectively, and the factor $(1 - \theta_{A})$ assures that $A$ constitutes a surface phase. The last term in Eq.~(\ref{eq:mc1}) accounts for the desorption of $c$ from the growth surface, and $\sigma$ (with the dimension of $\text{nm}^2$) represents the cross section of colliding $c$ with O \cite{vogt2017a}. The impinging fluxes of $c$ and $r$ are denoted as $\phi_{c}$ and \fr, respectively. We note that the structure of the model introduced here is similar to the MEXCAT model given in Ref.~\cite{vogt2017a}. The improvement of the model given in this work and its generalization to elemental and molecular catalysts arises by taking a cationic-like, metastable adlayer $A$ into account. This allows MOCATAXY to be described for the growth of ternary systems involving molecular catalysts [e.\,g., \iso\ (Fig.~\ref{fig:cat}) and \sso\ (Fig.~\ref{fig:cat})] as well as elemental catalysts [e.\,g., In (Fig.~\ref{fig:catsup}) and Sn (Fig.~\ref{fig:catsup})].
\subsection{B.~Suboxide catalysts}
\noindent
Our results thus indicate that MOCATAXY takes place with a suboxide catalyst (e.g., for $A$ being \iso--O or \sso--O) and \textit{not} with an elemental catalyst (e.g., for $A$ being In--O or Sn--O). For conventional MBE, i.e., when using elemental source materials, we assume the reaction of the metal to form the suboxide (e.g., $2\text{Ga} + \text{O} \rightarrow \text{\gso} = r$) occurs very rapidly \cite{vogt2018b}, and thus, the catalysis takes place between the suboxide reactant $r$ and $A$, satisfying reactions (\ref{eq:c1}) and (\ref{eq:c2}). Should the reaction of $2\text{Ga} + \text{O} \rightarrow \text{\gso}$ not occur very rapidly, we would see Ga desorption (at least a fraction of it), and a plateau in \gr\ in the adsorption-controlled regime during the
growth of \gao\ by conventional MBE; similar to the \gr\ plateau, e.g., observed during the growth of GaN by conventional MBE \cite{garrido2008}. The subsequent desorption of the rapidly formed \gso\ is the growth-rate-limiting step in conventional MBE as it removes active O from the growth front (by forming \gso), leading to the decrease in \gr\ in the adsorption-controlled regime as well as at elevated \tg\ \cite{tsai2010,vogtdiss,vogt2018b}. In $S$-MBE, the formation of \gso\ (through $2\text{Ga} + \text{O} \rightarrow \text{\gso}$) is bypassed since \gso\ is directly provided from the source. Therefore, the O-consuming step is now avoided and a plateau in \gr\ of \gao\ in the adsorption-controlled regime occurs \cite{vogt2021}. The assumption that MOCATAXY takes place with a suboxide catalyst is further supported by our finding in this work that the oxidation efficiency $\eta$ for \iso\ and \gso\ during MOCATAXY by $S$-MBE follows the same order in $\eta$ as for Ga and In during conventional MBE, i.e., $\eta_{\text{\iso}} = 2.8\eta_{\text{\gso}}$ for $S$-MBE [\text{this work}] versus $\eta_{\text{In}} = 2.8\eta_{\text{Ga}}$ for conventional MBE \cite{vogtdiss,vogt2017a}. Nevertheless, we emphasize that the microscopic origin and reaction pathways of MOCATAXY during $S$-MBE and conventional MBE requires further investigation to fully understand which species are indeed involved leading to the observed catalysis given in this work and presented in Refs.~\cite{vogtdiss,vogt2017a,kracht2017,vogt2018a,mazzo2019,mauze2020,mauze2020a,kneiss2019,wang2020}.

Solving Eqs.~(\ref{eq:mc1}) and (\ref{eq:mc2}) with respect to $n_c$ and $\theta_A$ yields $\text{\fc} = \gamma_c n_c$, consistent with our observation of negligible incorporation ($\approx 3\,\%$) of $c$ into the grown thin films for the data plotted in Fig.~\ref{fig:cat}. Inserting the solution for $\theta_A$ into Eq.~(\ref{eq:mc3}) yields the following expression, valid for first-order kinetics with $x = 1$,
\begin{equation}
\label{eq:cmod}
\text{\gr} = \frac{\alpha \text{\fc} \text{\fr} \text{\fo}}{\alpha J \text{\fr} + \text{\fc} \text{\fo}} \;.
\end{equation}
The free parameters are the pseudo-flux
\begin{equation}
\label{eq:pflux}
J = \frac{\gamma_c}{\sigma} = J_0 \, \text{exp}\left(-\frac{\Delta}{k_{\text{B}} \text{\tg}}\right)  
\end{equation}
with $J_0 = 1 \times 10^{14} \, \text{nm}^{-2}\, \text{s}^{-1}$ (assumed for all species as a first approximation \cite{lombardo1991}) and energy $\Delta$. The value of $\Delta$ depends linearly on $\phi_c$ (e.g., \fiso\ and \fsso),
\begin{equation}
\label{eq:delta}
\Delta\bigl(\phi_c\bigr) = \Delta_0 + \delta \phi_c \,,
\end{equation}
with $\Delta_0$ denoting the evaporation enthalpy, e.g, of \iso\ and \sso, and $\delta$ describing its increase with increasing $\phi_c$. 
The other free parameter is the catalytic activity coefficient, $\alpha$. For $\alpha$ (ranging from 1 to 0), we use a linear approximation depending on impinging \fo, i.e.,
\begin{equation}
\label{eq:alpha}
\alpha\bigl(\text{\fo}\bigr) = 1 - b \text{\fo} \,,
\end{equation}
with $b$ describing the decrease of $\alpha$ with \fo. By an iterative approach of our model to the experimental growth rate data, the values obtained of $\Delta_0$, $\delta$, and $b$ for \iso\ and \sso\ are given in Table \ref{tab2}.
\begin{table}[t]
  \centering	
    \begin{tabular}{||c|c|c|c||}
		 \toprule
   &$\; \Delta_0 \, (\text{eV}$)\; &$\; \delta \, (\text{meV}\,\text{nm}^2\,\text{s}) \;$ &$\; b \, (\text{nm}^2\,\text{s}$)\;  \\ \midrule
	 $\; \upbeta$-\gao($\bar{2}$01):\iso \;  & $2.6 \pm 0.05  $ \cite{val1977}    & $ 30 \pm 2 $    & $0.07 \pm 0.01$    \\ \hline
	 $\; \upbeta$-\gao(010):\iso \;          & $2.6 \pm 0.05  $ \cite{val1977}    & $ 30 \pm 2 $    & $0.07 \pm 0.01$    \\ \hline
	 $\; \upbeta$-\gao($\bar{2}$01):\sso \;  & $2.9 \pm 0.05  $ \cite{colin1965}  & $ 50 \pm 4 $    & $0.20 \pm 0.03$    \\ \hline
	 $\; \upbeta$-\gao(010):\sso \;          & $2.9 \pm 0.05  $ \cite{colin1965}  & $ 10 \pm 1 $    & $0.33 \pm 0.05$    \\ \hline
	              \ino(111):\sso \;          & $3.2 \pm 0.05  $                   & $ 400 \pm 10 $  & $0.33 \pm 0.05$    \\ \hline
\bottomrule
	\end{tabular} 
\caption{The values of $\Delta_0$, $\delta$, and $b$ used in Eqs.~(\ref{eq:delta}) and (\ref{eq:alpha}) for different catalytic systems and growth surfaces.}\label{tab2}
\end{table}	

The evaporation enthalpies $\Delta_0$ of \iso\ and \sso\ on \gao\ surfaces correspond to the values given in the literature, as noted in Table \ref{tab2}. Only the value of $\Delta_0$ for \sso\ on the \ino(111) surface is slightly above the literature value, indicating an additional energy term caused by vertical and lateral interactions between adsorbed species and the \ino(111) surface. The increase in $\Delta$ with $\phi_c$, described by the parameter $\delta$, indicates an increase in lateral adsorbate binding energy with increasing $\phi_c \propto n_c$. An analogous behavior has also been observed for the desorption kinetics of In on \gao\ \cite{vogt2017a} and Ga on GaN \cite{he2006}. We explain the linear decrease of $\alpha$ with \fo\ by a linear increase of O adsorbates with \fo, i.e., $\text{\fo} \propto n_{\text{o}}$. It has been shown that an increase in \fo\ promotes the incorporation of In during the growth of \ingao\ by conventional MBE; this decreases the catalytic strength of In for the formation of \gao\ \cite{vogt2016c,vogtdiss,vogt2017a,mazzo2020}. We therefore explain the decrease in $\alpha$ with increasing $\text{\fo} \propto n_{\text{o}}$ by a decreasing diffusion length of $c$ and $r$, and thus, a decrease in the probability of reaction (\ref{eq:c2}) to occur.
\subsection{C.~Elemental catalysts}
\noindent
We have applied the above model to the data published in Refs.~\cite{vogt2017a} and \cite{kracht2017} for MOCATAXY during conventional MBE growth of \gao, i.e., using elemental In and Sn as catalysts (see Appendix). For the former, our model and the parameters given in this Letter can describe the data published in Ref.~\cite{vogt2017a} with the same accuracy. For the latter, we provide the first mathematical description of a catalytic system involving the element Sn (see Appendix). These results, and the comparison among different MBE variants (i.e., $S$-MBE and conventional MBE), confirm the accuracy of our refined and generalized MEXCAT model, describing MOCATAXY for elemental and molecular catalysts.
\subsection{D.~Obtained `pseudo' \gao/\alo\ heterostructure}
\noindent
\begin{figure}[t]
\includegraphics[scale=0.645]{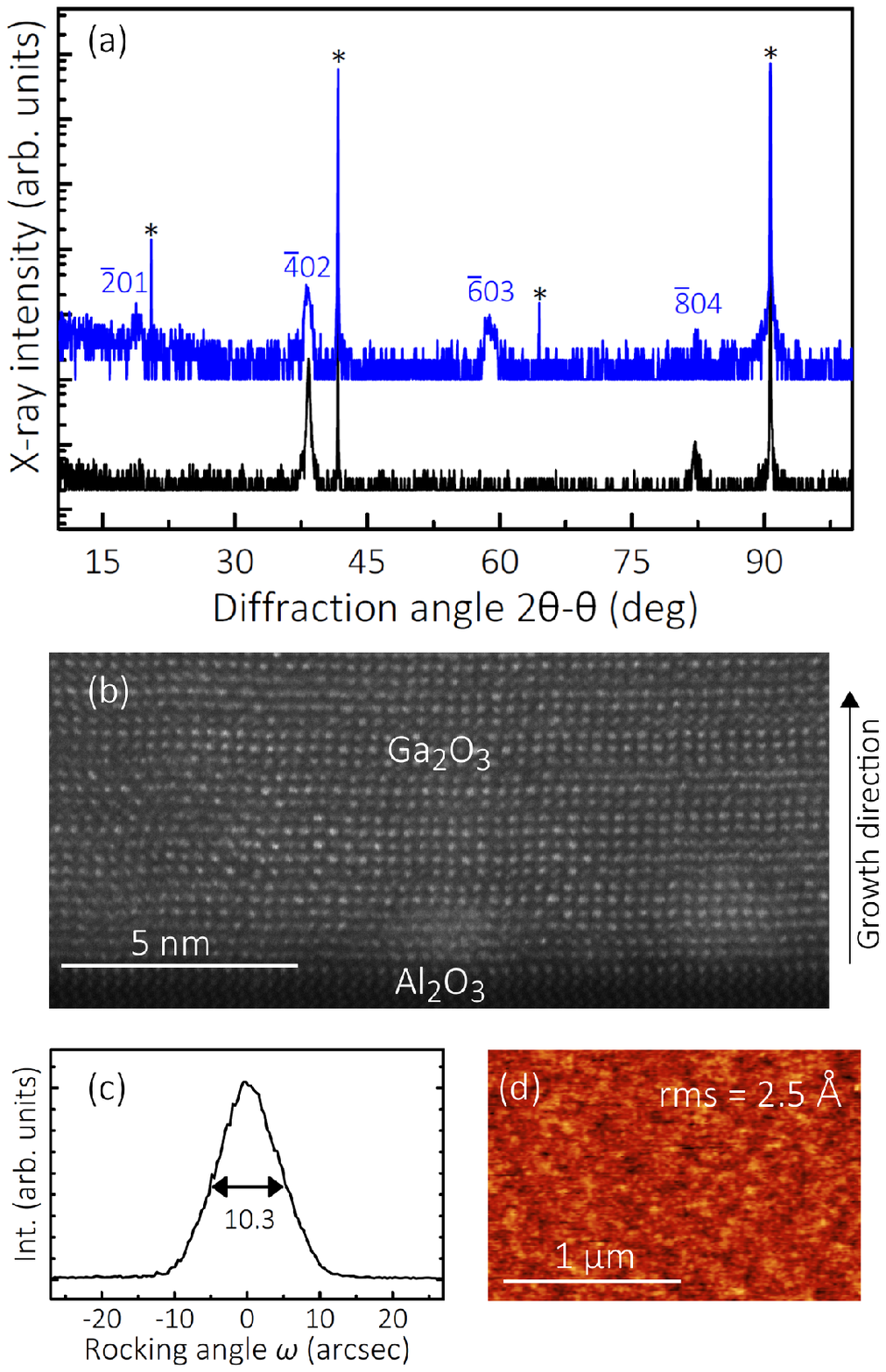}
\caption{(a) Longitudinal XRD scans recorded from an \sso-catalyzed \gao\ film (black trace) and a non-catalyzed $\upbeta$-\gao($\bar{2}$01) film (blue trace) grown by $S$-MBE on \alo(0001) substrates. Reflections labeled by an asterisk originate from the substrate. The reflections from the \gao\ films are marked in the figure. (b) STEM image along the [100] zone axis of the \gao\ thin film. (c) Transverse XRD scan across the $\bar{4}02$ peak of the same \sso-catalyzed \gao\ film [black trace in (a)], with its full width at half maximum indicated in the figure. (d) Surface morphology of the same film \sso-catalyzed \gao\ film; it has a root means square roughness of $2.5\,\text{\AA}$. The data depicted in (a) [black trace], (b), (c), and (d) were taken from the same \gao\ film plotted by the solid star in Fig.~\ref{fig:cat}(f).}
\label{fig:xrd}
\end{figure} 
As a result of combining $S$-MBE with MOCATAXY, in Figs.~\ref{fig:xrd}(a)--\ref{fig:xrd}(d), we demonstrate an unprecedented `pseudo' $\upbeta$-\gao($\bar{2}$01)/\alo(0001) heterostructure with unparalleled crystalline perfection. Figure \ref{fig:xrd}(a) shows $\theta$-$2\theta$ x-ray diffraction (XRD) scans of a $90\,\text{nm}$ thick \sso-catalyzed \gao\ film (black trace) and a $40\,\text{nm}$ thick non-catalyzed \gao\ film grown by $S$-MBE under similar growth conditions ($\text{\tg} = 500\text{\utg}$, $\text{\fgso} = 0.9\text{\pfu}$, and $\text{\fo} = 2.3\text{\pfu}$). The SnO-catalyzed \gao\ film is the same film that is plotted in Fig.~\ref{fig:cat}(f) as a solid-black star. It was grown at $\text{\tg} = 690\text{\utg}$, $\text{\fgso} = 1.3\text{\pfu}$, and $\text{\fo} = 2.3\text{\pfu}$. The reflections of the non-catalyzed \gao\ thin film coincide with the $\upbeta$-\gao($\bar{2}$01) phase grown with its ($\bar{2}$01) plane parallel to the (0001) plane of the \alo\ substrate. In contrast, the XRD scan of the \sso-catalyzed \gao\ film shows only the even reflections of the $\upbeta$-\gao\ phase grown with its ($\bar{2}$01) plane parallel to the (0001) plane of the \alo\ substrate. We speculate that the `pseudo' $\upbeta$-\gao($\bar{2}$01)/\alo(0001) contains an aperiodic occurrence of low-energy stacking faults, parallel to the ($\bar{2}$01) plane of the \gao\ film that reduce the number of the observed XRD diffraction peaks of the `pseudo' $\upbeta$-\gao($\bar{2}$01) phase. This hypothesis is consistent with calculations showing that the stacking fault energy in $\upbeta$-\gao\ is low \cite{fu2019,mccluskey2020}, experimental observation of high densities of stacking faults in $\upbeta$-\gao\ films \cite{schewski2016}, and the scanning transmission electron microscopy (STEM) image depicted in Fig.~\ref{fig:xrd}(b). The epilayer shows a single-crystalline structure with an abrupt interface to the \alo(0001) substrate at the bottom of the image. Figure \ref{fig:xrd}(c) shows a transverse scan (rocking curve) across the symmetric $\bar{4}02$ reflection of the same `pseudo' $\upbeta$-\gao($\bar{2}$01) film. The measured full width at half maximum (FWHM) is just $\Delta \omega \approx 10\,\text{arcsec}$; this is a measure of the out-of-plane mosaic spread of the thin film. Figure \ref{fig:xrd}(d) depicts an atomic force microscope (AFM) image of the surface morphology of the same SnO-catalyzed \gao\ film; it has a root mean square roughness (rms) of $2.5\,\text{\AA}$. The obtained rocking curve of $\Delta \omega \approx 10\,\text{arcsec}$ and smooth surface morphology of $\text{rms} = 2.5\,\text{\AA}$ provide the best results obtained for any \gao\ thin film grown on \alo(0001) by any method. We point out that optimizations utilizing the unprecedented growth regimes---becoming accessible by the combination of $S$-MBE with MOCATAXY---still need to be performed to further improve the crystalline perfection of the grown \gao-based heterostructures.

We note that in the absence of a catalyst, \gao\ does not form at this high $\text{\tg}$ of $690\text{\utg}$. In addition, at a lower catalyst flux of SnO [depicted as the open star in Fig.~\ref{fig:cat} (f)] compared with the sample grown at a higher SnO flux [depicted as the solid star in Fig.~\ref{fig:cat} (f) and Fig.~\ref{fig:xrd}] we have not measured the `pseudo' $\upbeta$-\gao($\bar{2}$01) phase, but instead see the `conventional' $\upbeta$-\gao($\bar{2}$01) peaks by XRD (data not shown in this work). We therefore conclude that extending the kinetic (e.g., higher possible \tg) and thermodynamic limits (e.g., different surface chemical potential) by combining $S$-MBE with MOCATAXY benefits the formation of the `pseudo' $\upbeta$-\gao($\bar{2}$01)/\alo(0001) heterostructure that has unparalleled crystalline perfection.
\section{IV.~Conclusion}
\noindent
As we have demonstrated, the nature of the model derived to describe MOCATAXY using elemental and molecular catalysts does $not$ depend on the specific growth surface. We note, however, that the growth surface may change the kinetic parameters used in our model as shown for the examples of $\upbeta$-\gao($\bar{2}$01) and $\upbeta$-\gao(010) using \iso\ or \sso\ as catalysts in this work.

Finally, the increase in the growth rates of \gao\ and \ino\ by $S$-MBE that occur when using the catalysts \sso\ and \iso, demonstrates MOCATAXY as a potentially inherent feature in conventional MBE growth \cite{vogt2017a,kracht2017,vogt2018a,mazzo2019,mauze2020,mauze2020a} as well as in $S$-MBE growth. Furthermore, our results provide deeper insight into this catalysis, indicating MOCATAXY occurs through a suboxide catalyst rather than with an elemental catalyst. This more broad applicability of MOCATAXY opens an unprecedented path for the epitaxial synthesis of thin films by intentionally extending the kinetic and thermodynamic limits during their growth processes. If successful, this could enable the growth of (yet) unknown crystal phases and unprecedented functional electronic materials.
\section{V.~Appendix}
\noindent
We apply the generalized MOCATAXY model, Eqs.~(\ref{eq:mc1})--(\ref{eq:alpha}), to published catalytic data using elemental In \cite{vogt2017a} and elemental Sn \cite{kracht2017} as catalysts for the growth of \gao\ by MBE. The parameters used in Eqs.~(\ref{eq:delta}) for In and Sn are collected in Table \ref{tab3}. Within experimental uncertainty, the obtained $\Delta_0$ corresponds to the vaporization enthalpies of elemental In \cite{alcock1984} and Sn \cite{colin1965} as given in the literature; the values are given in Table \ref{tab3}. Moreover, for our model we use $\alpha = 1$ (the catalytic activity), the same value of $\alpha$ as used in Ref.~\cite{vogt2017a}. 

\begin{figure}[t]
\includegraphics[scale=0.975]{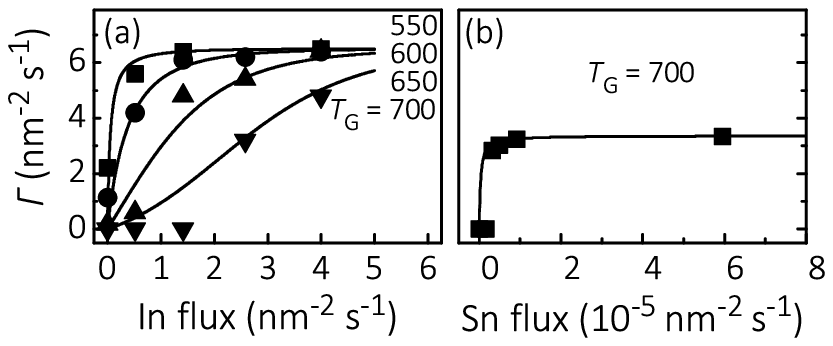}
\caption{(a) Growth rate (\gr) of \gao\ as a function of the In flux at different growth temperatures \tg\ and an active atomic oxygen flux \fo = 19.2\pfu. Data are taken from Ref.~\cite{vogt2017a}. (b) \gr\ of \gao\ as a function of the Sn flux, with \fo\ = 7.5\pfu\ used for the model. Data are taken from Ref.~\cite{kracht2017}. Model predictions by Eqs.~(\ref{eq:mc1})--(\ref{eq:alpha}) are drawn as solid lines using the kinetic parameters given in Table \ref{tab3}. The unit of the \tg\ values given is \utg.}
\label{fig:catsup}
\end{figure}

Figures \ref{fig:catsup}(a) and \ref{fig:catsup}(b) show our results for the growth of \gao\ by MBE when supplying In and Sn as elemental catalysts, respectively. Our model drawn in Fig.~\ref{fig:catsup}(a) (solid lines) describes the data published in Ref.~\cite{vogt2017a} with the same accuracy as the previously established metal-exchange catalysis (MEXCAT) model \cite{vogt2017a}. Figure \ref{fig:catsup}(b) plots the catalyzed growth rate data of \gao\ taken from Ref.~\cite{kracht2017}. Our model applied to this data is drawn as the solid line. It precisely describes the Sn-catalyzed growth of \gao\ by MBE and is the first quantitative description of MOCATAXY using Sn as a catalyst. To model the Sn-catalyzed data, we linearly extrapolated the value of $\phi_{\text{O}}^{\text{Sn}}$ given in Eq.~(\ref{eq:s11}) while using the O flux conditions published in Ref.~\cite{kracht2017}.

To conclude, our model introduced in the main text is able to the describe the growth of \gao\ and \ino\ by $S$-MBE as well as conventional MBE when using elemental and molecular catalysts. This is achieved by introducing a cationic-like, catalytic adlayer $A$.
\subsection{A.~Detailed model of the growth of III-VI materials by $S$-MBE}
\noindent
The analytical solution of the growth rate, \gr, of Eqs.~(\ref{eq:m1})--(\ref{eq:m3}) for III-VI compound materials is:
\begin{equation} 
\label{eq:s1}
\begin{aligned}
\text{\gr} &= \Biggl(\zeta_{-}^2  P^2 + \xi_{-}^{\frac{2}{3}} - P \Bigl(3 - 2  \zeta_{+}  \xi_{-}^{\frac{1}{3}}  \Bigr)\Biggr) \; \times \\
           &  \Biggl(\zeta_{-}^2  P^2 + \xi_{-}^{\frac{2}{3}} - P \Bigl(3 - \zeta_{-}  \xi_{-}^{\frac{1}{3}}  \Bigr)\Biggr) \times \Biggl(- 54  P  \xi_{+} \Biggr)^{-1} 
\end{aligned}
\end{equation}
with
\begin{align}
\zeta_{+} &= 2\text{\fr} - \varsigma \text{\fo} \;, \\ 
\zeta_{-} &= - 2\text{\fr} + \varsigma \text{\fo} \;, \\ 
\xi_{+}   &= \Bigl(\zeta_{+}  P - 9  P  \psi - \sqrt{27 \Psi}\Bigr)^3 \;, \\
\xi_{-}   &= - \Bigl(\zeta_{+}  P + 9  P  \psi + \sqrt{27 \Psi}\Bigr)^3 \;, \\
\psi        &= \text{\fr} + \varsigma \text{\fo} \;,
\end{align}
and
\begin{equation}
\label{eq:s7}
\begin{aligned}
\Psi        &= P^3 \Biggl(1 + \zeta_{-}^3  P^2  \varsigma  \text{\fo} +  \\
            & + P \; \times \Bigl(-\text{\fr}^2 + 10  \varsigma  \text{\fr} \text{\fo} + 2 \varsigma^2  \text{\fo}^2 \Bigr)\Biggr) \;.
\end{aligned}
\end{equation}
The growth rates of \gao\ and \ino\ as presented in Fig.~\ref{fig:ino} are explicitly modeled with Eqs.~(\ref{eq:s1})--(\ref{eq:s7}).
\subsection{B.~Detailed model of the growth IV-VI materials by $S$-MBE}
\noindent
The analytical solution for \gr\ of Eqs.~(\ref{eq:m1})--(\ref{eq:m3}) for IV-VI compound materials is:
\begin{equation}
\label{eq:s8}
\begin{aligned}
\text{\gr} &= \frac{1 + P  \psi_{+} - \sqrt{1 + 2 P \psi_{+} + P^2 \psi_{-}^2 }}{2 P}
\end{aligned}
\end{equation}
with
\begin{equation}
\label{eq:s9}
\psi_{\pm}     = \text{\fr} \pm \varsigma \text{\fo} \;.
\end{equation}
We predict that \gr\ of IV-VI materials (e.g., \sno)---obtained by $S$-MBE---may be modeled by Eqs.~(\ref{eq:s8}) and (\ref{eq:s9}) with the same accuracy as \gr\ modeled for III-VI materials (e.g., \gao\ and \ino) by Eqs.~(\ref{eq:s1})--(\ref{eq:s7}); this is demonstrated in the main text.
\subsection{C.~Conversion factors for \fr\ and \fo}
\noindent
The model uses reactant fluxes, \fr\ (e.g., \gso, \iso, \sso, Ga, In, Sn), and active atomic oxygen fluxes, \fo\ (from active O$_3$ or O species), in $\text{nm}^{-2}\,\text{s}^{-1}$. 
\\[1em]
The conversion factors for $S$-MBE used in an ozone MBE system are:
\begin{equation}
\begin{aligned}
\phi_{\text{\gso}}^{\text{QCM}} &= 1 \, \text{\AA} \, \text{s}^{-1} \, \rightarrow \, \phi_{\text{\gso}} = 0.38\text{\pfu}  \\
\phi_{\text{\iso}}^{\text{QCM}} &= 1 \, \text{\AA} \, \text{s}^{-1} \, \rightarrow \, \phi_{\text{\iso}} = 0.25\text{\pfu}  \\
\phi_{\text{\sso}}^{\text{QCM}} &= 1 \, \text{\AA} \, \text{s}^{-1} \, \rightarrow \, \phi_{\text{\sso}} = 0.45\text{\pfu}  
\end{aligned}
\end{equation}
\begin{equation}
\begin{aligned}
\label{eq:fc}
P_{\text{O}}^{\text{\gso}} &= 1 \times \, 10^{-6} \, \text{Torr} \, \rightarrow \, \phi_{\text{O}}^{\text{\gso}} = 0.73\text{\pfu}  \\
P_{\text{O}}^{\text{\iso}} &= 1 \times \, 10^{-6} \, \text{Torr} \, \rightarrow \, \phi_{\text{O}}^{\text{\iso}} = 2.05\text{\pfu}  \\
P_{\text{O}}^{\text{\sso}} &= 1 \times \, 10^{-6} \, \text{Torr} \, \rightarrow \, \phi_{\text{O}}^{\text{\sso}} = 2.25\text{\pfu}  
\end{aligned}
\end{equation}
The reactant fluxes, $\phi_r^{\text{QCM}}$, are measured prior to growth by a quartz crystal microbalance (QCM) in $\text{\AA}\,\text{s}^{-1}$, with the density of the QCM set to $1 \, \text{g}\,\text{cm}^{-3}$. These QCM readings are readily converted to absolute fluxes using the known masses of the \gso, \iso, and \sso\ molecules that condense onto the QCM. The active atomic oxygen fluxes given result from an oxidant, $P_{\text{O}}^r$, with mixtures of O$_2$ and $\sim$ 10\% O$_3$ in the oxygen molecular-beam \cite{theis1997}. The different oxidation efficiencies were taken from Ref.~\cite{vogtdiss} while using the results obtained for $S$-MBE.
\begin{table}[t]
  \centering	
    \begin{tabular}{||c|c|c||}
		 \toprule
   &$\; \Delta_0 \, (\text{eV}$)\; &$\; \delta \, (\text{meV}\,\text{nm}^2\,\text{s}) \;$ \\ \midrule
	 \gao:In \;    & $2.42 \pm 0.05  $ \cite{alcock1984}    & $ 41\pm 2 $      \\ \hline
	 \gao:Sn \;    & $2.91 \pm 0.07  $ \cite{colin1965}    & $ 10 \pm 2 $      \\ \hline
\bottomrule
	\end{tabular} 
\caption{Values of $\Delta_0$ and $\delta$ used in Eq.~(14) (main text) for the catalytic model when using In and Sn as catalysts.}\label{tab3}
\end{table}	
\\[1em]
The conversion factors for conventional MBE used in an oxygen plasma-assisted MBE system are:
\begin{equation}
\begin{aligned}
P_{\text{Ga}}^{\text{BEP}}   &= 1 \times  \, 10^{-7} \, \text{Torr} \, \rightarrow \, \phi_{\text{Ga}} = 1.2\text{\pfu}   \\
P_{\text{In}}^{\text{BEP}}   &= 1 \times  \, 10^{-7} \, \text{Torr} \, \rightarrow \, \phi_{\text{In}} = 0.6\text{\pfu}  \\
P_{\text{Sn}}^{\text{BEP}}   &= 1 \times  \, 10^{-7} \, \text{Torr} \, \rightarrow \, \phi_{\text{Sn}} = 1.5\text{\pfu}  
\end{aligned}
\end{equation}
\begin{equation}
\begin{aligned}
\label{eq:s11}
\phi_{\text{O}}^{\text{Ga}} &= 1 \, \text{SCCM} \, \rightarrow \, \phi_{\text{O}}^{\text{Ga}} = 9.8\text{\pfu}  \\
\phi_{\text{O}}^{\text{In}} &= 1 \, \text{SCCM} \, \rightarrow \, \phi_{\text{O}}^{\text{In}} = 27.4\text{\pfu}  \\
\phi_{\text{O}}^{\text{Sn}} &= 1 \, \text{SCCM} \, \rightarrow \, \phi_{\text{O}}^{\text{Sn}} = 30.1\text{\pfu}   
\end{aligned}
\end{equation}
The beam-equivalent pressure (BEP), $P_r^{\text{BEP}}$, of the reactant fluxes is measured prior to growth by an ion gauge in $10^{-7}\,\text{Torr}$. $P_r^{\text{BEP}}$ is converted from a pressure into a flux using the kinetic theory of gases \cite{schlombook}. The oxygen flux, $\phi_{\text{O}}^r$, was measured in standard cubic centimeters per minute (SCCM), and a radio-frequency plasma power of $300\,\text{W}$ was applied.
The conversion factors were taken from Ref.~\cite{vogtdiss} and applied to the data given in Ref.~\cite{kracht2017}. We note the conversion factors \cite{vogtdiss} and data \cite{kracht2017} were obtained in different MBE systems. Thus, the actual conversion may differ slightly due to the different geometries of the MBE systems used. Nevertheless, using the same conversion factors for different MBE systems is a practical and reasonable approach.
\section{IV.~ACKNOWLEDGMENTS}
\noindent
K.A., J.P.M., D.J., H.G.X., D.A.M., and D.G.S. acknowledge support
from the AFOSR/AFRL ACCESS Center of Excellence under
Award No. FA9550-18-1-0529. J.P.M. also acknowledges support
from the National Science Foundation within a Graduate Research
Fellowship under Grant No. DGE-1650441. P.V. and J.P. acknowledges support
from ASCENT, one of six centers in JUMP, a Semiconductor
Research Corporation (SRC) program sponsored by DARPA. 
F.V.E.H. acknowledges support from the Alexander von Humboldt
Foundation in the form of a Feodor Lynen fellowship. F.V.E.H.
also acknowledges support from the National Science Foundation
(NSF) [Platform for the Accelerated Realization, Analysis and
Discovery of Interface Materials (PARADIM)] under Cooperative
Agreement No. DMR-2039380. J.P.M. acknowledges the support of a National Science 
Foundation Graduate Research Fellowship under Grant No.~DGE-1650441. This work made use of the
Cornell Center for Materials Research (CCMR) Shared Facilities,
which are supported through the NSF MRSEC Program (Grant No.
DMR-1719875). Substrate preparation was performed, in part, at the
Cornell NanoScale Facility, a member of the National Nanotechnology
Coordinated Infrastructure (NNCI), which is supported by
the NSF (Grant No. NNCI-2025233).

\bibliography{references}
\end{document}